\documentclass[12pt]{article}
\usepackage{epsfig}
\begin{document}

\markboth{}{\it Expanding Space}

\title{ Flat, Uniformly Expanding Space in five Exercises}

\author{Bernd A.\ Berg \\ Department of Physics\\ Florida State University,
Tallahassee, Florida 32306}
% \email{berg@hep.fsu.edu} % \date{April 21, 2023} 
\date{\today}
\bigskip

\maketitle %\narrowtext

\centerline{\bf Abstract} \bigskip
Exercises with solutions are presented which should allow advanced 
undergraduate students to understand properties of a flat, uniformly 
expanding space. No knowledge of general or special relativity is 
needed besides that the speed of light $c$ is a constant. The material 
could be used stand alone or would fit well at the end of a treatment 
of special relativity, 

Using a reasonable value for the Hubble constant $H$, trajectories of 
light are calculated within the model, followed by a calculation of the 
Hubble horizon, and the synchronization of clocks is demonstrated.
\bigskip

{\bf Introduction} \bigskip

We consider flat space $(\vec{x},t)$, where a common time is defined 
for all space points (synchronization of clocks can be achieved and is 
discussed in exercise 5). Transformations to other coordinate systems
are not considered (the model is not invariant under general coordinate
transformations). Let us focus on two space points which are 
initially, i.e., at common time $t=0$, located a distance $x_0>0$ 
apart. The distance between them is assumed to increase according to 
\begin{eqnarray} \label{dt}
  x_1(t) = x_0 \exp(tH)\, 
\end{eqnarray}
due to the Hubble expansion of space. Here $H$ is the Hubble constant, 
which is a constant in space, and  we take it also to be a constant 
in time. For sufficiently large time $t$ the two points separate
faster than with the speed of light, which is often an issue of 
confusion for students who just learned special relativity. Here 
the resolution is that $c\,dt$, with $c$ constant is only one of 
two contributions to the infinitesimal motion of light.

It has to  pointed out that our exercises aim at a qualitative 
understanding, and we cannot expect quantitatively relevant estimates 
of physical observables. Nevertheless we shall in numerical calculations 
use physical values for $c$ and $H$. Estimates for $H$ are in the range 
\cite{NASA} from 68 to 74$\ km/s$ per $Mpc$. We use 
\begin{eqnarray} \label{H}
  H = 71\ \rm km/s\ per\ Mpc\ .
\end{eqnarray}
In the following
we take the $x$-axis along the straight line which connects the two 
points. The location of our first space point is by definition the 
zero of the axis, $x=0$ for all $t$.  Our second space point lies at 
$x_1(t)$, given by Eq.(\ref{dt}), on the $x$-axis. At position $x_0=x_1
(0)$ we imagine a laser which emits at $t=0$ an ultrashort photon 
ray towards the first space point at $x=0$, where a photon detector is 
located. There are no gravitational interactions as we are dealing
with a model of pure space.
\bigskip

{\bf 2. Exercises and Solutions}
\bigskip

{\bf Exercise 1: The Model.} \medskip

Consider the infinitesimal motion $dx$ of an ultrashort photon ray that 
is at time $t=0$ emitted at space point two towards space point one.
Find and verify the analytical solution for the travel path $x(t)$ 
\medskip

{\bf Solution:} \medskip

For the position of the photon ray we derive a differential equation
from the infinitesimal motion:
\begin{eqnarray} \nonumber
  dx &=& -c\,dt + x\,\{\exp[H\,(t+dt)] - \exp[H\,t] \}\ 
    \ =\ -c\,dt + x\,\exp[H\,t]\,\{\exp[H\,dt] - 1\}\\ \label{dx}
     &=& -c\,dt + x\,\exp[H\,t]\,H\,dt\ \Rightarrow\ 
  \frac{dx}{dt}\ =\ -c + x\,H\,\exp[tH] \ ,
\end{eqnarray}
where $c$ is the speed of light, and $H$ the Hubble constant. The 
solution is (Mathematica)
\begin{eqnarray} \label{xt}
  x(t) = \exp\left(e^{tH}\right)\,\left( K_1 - \frac{c}{H}\,
         Ei\left(-e^{tH}\right)\right)\ ,
\end{eqnarray}
where $K_1$ is an integration constant, and  $Ei$ is the exponential 
integral
\begin{eqnarray} \label{Ei}
  Ei(y) = \int_{-\infty}^y dt'\,\frac{e^{t'}}{t'} = \gamma + \ln(|y|)
  - \sum_{k=1}^{\infty} \frac{y^k}{k\,k!}\ .
\end{eqnarray}
Here $\gamma=\lim_{n\to\infty}\left(-\ln(n)+\sum_{k=1}^n 1/k\right)$ 
is the Euler-Mascheroni constant. In the following the solution 
(\ref{xt}) is verified by differentiation. It is convenient to introduce 
the variable $y=e^{tH}>0$ and to break Eq.(\ref{dx}) up into
\begin{eqnarray} \label{dxdydydt}
  \frac{dx}{dt}\ = \frac{dx}{dy}\frac{dy}{dt}\,,\ \ {\rm where}\
  \ \frac{dy}{dt} = H\,e^{tH} = H\,y\ \ {\rm holds}.
\end{eqnarray}
It remains to calculate $dx/dy$. For $x(y)$ we get from Eq.(\ref{xt})
\begin{eqnarray} \label{xy}
  x(y) = e^y\left(K_1 - \frac{c}{H}\,Ei(-y)\right)\ .
\end{eqnarray}
The $y$ derivative of the exponential integral $Ei(-y)$ is
\begin{eqnarray} \label{dEidy}
  \frac{dEi(-y)}{dy} = - \frac{dEi(-y)}{d(-y)} = - \frac{d\ }{d(-y)} 
  \int_{-\infty}^{-y} dt'\,\frac{e^{t'}}{t'} = - \frac{e^{-y}}{-y}
  = \frac{e^{-y}}{y} \ .
\end{eqnarray}
So we get
\begin{eqnarray} \label{dxdy}
  \frac{dx}{dy} = e^y\left(K_1 - \frac{c}{H}\,Ei(-y)\,\right)
                - \frac{c}{H\,y}\ .
\end{eqnarray}
Multiplying with $dy/dt=H\,y$ yields
\begin{eqnarray} 
  \frac{dx}{dy}\frac{dy}{dt} = e^y \left(K_1 - \frac{c}{H}\,Ei(-y)
                               \right)\,H\,y - c\ .
\end{eqnarray}
Inserting $y=e^{tH}$ gives 
\begin{eqnarray} 
  \frac{dx}{dt} = \exp\left(e^{tH}\right)\,\left(K_1 - \frac{c}{H}\,
  Ei(-e^{tH}) \right)\,H\,\exp(tH) - c = x\,H\,\exp(tH) - c\,,
\end{eqnarray}
where Eq.(\ref{xt}) has been used backwards for the last equal sign.
\bigskip % \vfil\eject

{\bf Exercise 2: Trajectories of Light.} \medskip

\begin{figure}[tb] \vspace{7pc}
\centerline{\hbox{ \psfig{figure=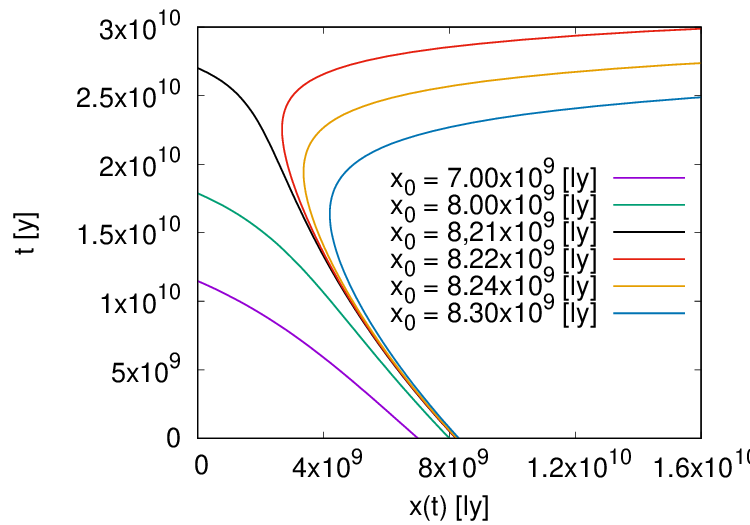,width=14cm} }}
\caption{Trajectories of light for $H=$ 71 km/s per Mpc and 
various initial distances $x_0$.} \label{fig_xt} \end{figure}

Use numerical values for $c,\ H$ and $x_0$ to obtain the $x(t)$ 
trajectories of Fig.(\ref{fig_xt}), where the units are years $[y]$ 
and light years $[ly]$. 
\medskip \eject

{\bf Solution:} \medskip

The integration constant $K_1$ in Eq.(\ref{xt}) is determined by the 
initial condition at time $t=0$, $x(0)=x_0>0$:
\begin{eqnarray} \label{K1}
  x_0 = e^1\left(K_1-\frac{c}{H}\,Ei(-1)\right)\ \Rightarrow\ K_1 =
  \frac{x_0}{e^1} + \frac{c}{H}\,Ei(-1)\,,\ Ei(-1)=-0.219384\dots\ .
\end{eqnarray}
Inserting $K_1$, Eq.(\ref{xt}) becomes
\begin{eqnarray} \label{xt1}
  x(t) = \exp\left(e^{tH}\right)\,\left(\frac{x_0}{e^1} +
  \frac{c}{H}\,Ei(-1) - \frac{c}{H}\,Ei\left(-e^{tH}\right)\,
  \right)\ .
\end{eqnarray}
To plot $x(t)$ we have to insert numbers for $c$ and $H$, We measure 
the time in units of years $[y]$. A light year $[ly]$ is defined as 
one year times the speed of light $c$. Therefore, the speed of light 
is one in units of years and light years, 
\begin{eqnarray} \label{cy}
  c=1\ [ly]/[y]\ . 
\end{eqnarray}
Furthermore, we need to express $H$ of Eq.(\ref{H}) in units 
of inverse years $[1/y]$. One Mpc is $3.262\times 10^6$ [$ly$]. 
Converting $[ly]$ to $[km]$, the distance unit $[km]$ 
drops out of Eq.(\ref{H}), and the value of $H$ becomes
\begin{eqnarray} \label{Hy}
  H=2.30222\times 10^{-18}\ [1/s]=7.2603\times 10^{-11}\ [1/y]\ ,
\end{eqnarray}
where we have converted seconds $[s]$ to years $[y]$ in the last step. 
It is now straightforward to plot $x(t)$ of Eq.(\ref{xt1}) for  various
$x_0$ values, and so that the results agree with Fig.(\ref{fig_xt}).
\bigskip % \vfil\eject

\bigskip % \vfil\eject

{\bf Exercise 3: Hubble Horizon.} \medskip

Find the largest value, $x_0^{\max}$, of the initial distance $x_0$, 
so that the photon ray will reach its destination $x(t_a)=0$ for all 
initial values $x_0<x_0^{\max}$. We call $x_0^{\max}$ Hubble horizon 
and $t_a$ arrival time. Plot the relationship between initial condition 
$x_0$ and arrival times $t_a$.

In particular, consider $t_a =13.7\times 10^9\ [y]$, i.e., the present 
estimate \cite{AGE} of the age of the universe, and calculate the 
corresponding $x_0$. For these values indicate the vectors 
$(0,t_a)\to(x_0,ta)$ and $(x_0,t_a)\to(x_0,0)$ in the figure.
 What is the distance between point one and point two at time 
$t=t_a =13.7\times 10^9\ [y]$?
\medskip

{\bf Solution:} \medskip

In Fig.(\ref{fig_xt}) we see that trajectories with initial values 
$x_0\le 8.21\times 10^9\ [ly]$ reach their destination $x(t_a)=0$ at 
some arrival time $t_a$. To the contrary, for initial values $x_0\ge 
8.22\times 10^9\ [ly]$ the curves turn around before reaching $x=0$, 
and an arrival time $t_a$ does not exist. Hence, we have for the Hubble 
horizon $8.21\times 10^9\ [ly]<x_0^{\max}<8.22\times 10^9\ [ly]$.
To get an accurate number for $x_0^{\max}$ we set $t=t_a$ and $x(t_a)
=0$ in Eq.(\ref{xt1}), and find the following relation between $t_a$ 
and $x_0$:
\begin{eqnarray} \label{ta}
      Ei\left(-e^{t_aH}\right) = \frac{H\,x_0}{c\,e^1} + Ei(-1)\ .
\end{eqnarray}
The Hubble horizon is then obtained using the asymptotic behavior 
$\lim_{y\to\infty} Ei(-y) = 0$:
\begin{eqnarray} \label{x0max}
  x_0^{\max}=-\frac{c\,e^1\,Ei(-1)}{H}=8.2138\dots\times 10^9\ [ly]\ .
\end{eqnarray}
\begin{figure}[tb] \vspace{7pc}
\centerline{\hbox{ \psfig{figure=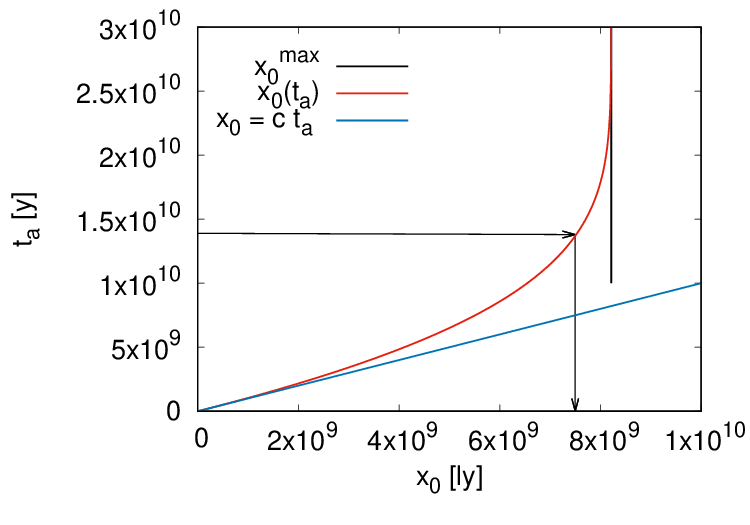,width=14cm} }}
\caption{Initial distance $x_0$ as function of the arrival time $t_a$,
where $x_0^{\max}$ is the Hubble horizon. The arrows point from $t_a 
=13.7\times 10^9\ [y]$ to $x_0 = 7.499\times 10^9\ [ly]$.} \label{fig_ta} 
\end{figure}
The relationship (\ref{ta}) between $x_0$ and $t_a$ is plotted 
in Fig.(\ref{fig_ta}). We see that for $t_a\to\infty$ the value 
$x_0^{\max}$ of Eq.(\ref{x0max}) is rapidly approached.

The value $t_a =13.7\times 10^9\ y$ corresponds to $x_0 = 7.499\times 
10^9\ [ly]$. See the arrows in Fig.(\ref{fig_ta}). During the travel 
time the distance between the two space points increases according 
to Eq.(\ref{H}): 
\begin{eqnarray} \label{xu}
  x_1(t_a) = x_0\,\exp(t_aH) =  20.473\ [ly]\ .
\end{eqnarray}
\bigskip \eject % \vfil\eject

{\bf Exercise 4: Leading order small $t$ expansion.} \medskip

Towards small $t_a$ it is seen in Fig.(\ref{fig_ta}) how the relation 
$x_0=c\,t$ becomes a good approximation. Let us discuss this quantitatively:
Expand $x(t)$ of Eq.(\ref{xt1}) to leading order in $t$. Transform the 
result into a first-order correction to the zero-order arrival time 
$t_a=x_0/c$, and compute the correction for $x_0=1\,[ly]$.
\medskip 

{\bf Solution:} \medskip 

To leading order in $t$ Eq.(\ref{xt1}) reads
\begin{eqnarray} \nonumber
  x(t)&=& \exp\left(1+tH\right)\,\left( \frac{x_0}{e^1} +
  \frac{c}{H}\,Ei(-1) - \frac{c}{H}\,Ei\left(-1-tH\right)\,\right) \\ 
  &=& e^1\,(1+tH)\,\left( \frac{x_0}{e^1} + \frac{c}{H}\,Ei(-1) - 
  \frac{c}{H}\,Ei\left(-1-tH\right)\,\right)\ . \label{xtapprox}
\end{eqnarray}
The derivative of the exponential integral is given by Eq.(\ref{dEidy}).
Defining $y=tH$ and using the result of Eq.(\ref{dEidy}) we get
\begin{eqnarray} 
  Ei(-1-y) = Ei(-1) - y\,\left.\frac{dEi(-y)}{d(-y)}\right|_{y=1}
           = Ei(-1) + tH\,e^{-1}\ .
\end{eqnarray}
The $Ei(-1)$ terms cancel out in Eq.(\ref{xtapprox}), and we arrive at 
\begin{eqnarray} \label{ct}
  x(t) = x_0\,(1+tH) - c\,t\
\end{eqnarray}
Transforming Eq.(\ref{ct}) into a correction to $t_a$, we get
\begin{eqnarray} \label{taapprox}
  t_a = \frac{x_0}{c} + \frac{x_0 H}{c^2}\ .
\end{eqnarray}
Assuming for $x_0$ the distance of $1\,[ly]$, we obtain from the
first term $1\,[y]$ (remember $c=1$ in our units). Using $H$ 
from Eq.(\ref{Hy}), we obtain from the second term the correction 
$7.260\time 10^{-11}\ [y]=2.290\time 10^{-3}\ [s]$. % 11Horizon.f
\bigskip % \vfil\eject

{\bf Exercise 5: Synchronization of clocks.} 
\medskip

In our model the time is globally defined. This requires a procedure to 
synchronize  clocks at different space positions. The situation is as 
follows: An observer A emits at time zero a photon ray in the direction 
of a second observer B. The photon ray is reflected by B and arrives at 
time $t_a>0$ back at A. The entire information at hand is $t_a$ from 
which A has to calculate the time at which B reflected the signal. Show  
that this is possible. 

Conclude this section with the numerical example $t_a=5 \times 10^9\ y$.
This is a value for which the $x_0(t_a)$ curve and the $x_0=c\,t_a$ line 
of Fig.(\ref{fig_ta}) have clearly separated. Draw a diagram of the
exchange of signals.
\medskip

{\bf Solution:} \medskip

We denote by $t^1_a$ the time the photon ray needs to travel from A to 
B, and by $t^2_a$ the time it needs to travel back from B to A. So we 
have
\begin{figure}[tb] \vspace{7pc}
\centerline{\hbox{ \psfig{figure=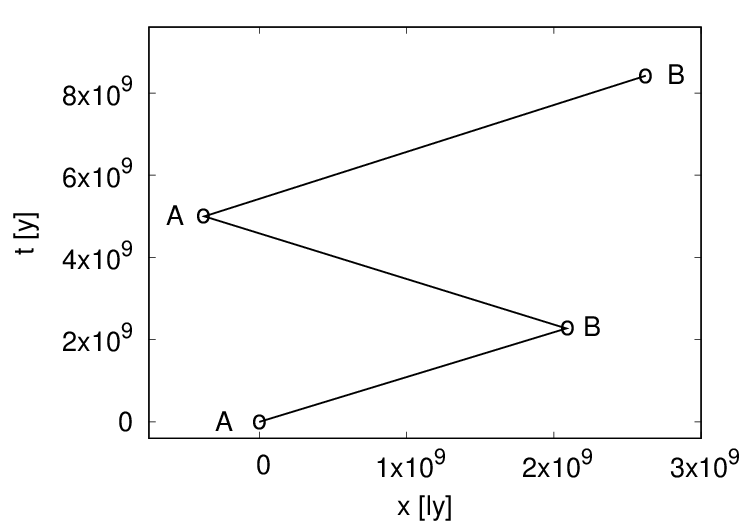,width=14cm} }}
\caption{Signal exchange A $\to$ B $\to$ A $\to$ B.} 
\label{fig_AB} \end{figure}
\bigskip % \vfil\eject
\begin{eqnarray} \label{ta1}
  t_a\ =\ t^1_a + t^2_a\ .
\end{eqnarray}
Let us start with the guess: $t_{g_1}=t^1_{g_1}+t^2_{g_1}$ with 
$t^1_{g_1}=t^2_{g_1}=t_a/2$. This is correct for $H=0$ the situation of 
special relativity. A would then tell B  to set its clock to $t^1_a =
t_a/2$ plus whatever time has elapsed on B's clock since receiving the 
signal (A and B are machines). 

However, for $H>0$ the guess is false because $t^1_{g_1}=t_a/2$ 
implies $t_{g_1}>t_a$. Namely, assume that the initial distance from 
A to B is $x^1_{g_1}$, then the subsequent distance from B to A is
according to Eq.(\ref{dt}) $x^2_{g_1}=x^1_{g_1} \exp(t^1_{g_1}\,H)
> x^1_{g^1}$. As long as $x^2_{g_1}<x_0^{\max}$ there is a return 
signal that needs a time $t^2_{g_1} > t^1_{g_1}$ to travel from B 
to A. See Fig.(\ref{fig_ta}). As $t^1_{g_1}=t_a/2$ it follows that 
$t_{g_1}=t^1_{g_1}+t^2_{g_1}>t_a$.

Now, with $t^1_{g_1}$ given we can calculate $t^2_{g_1}$ and, hence, 
$t_{g_1}$ explicitly. As $t_{g_1}(t^1_{g_1})$ is a monotonically 
increasing function, we can compute $t^1_a$ with the bisection method. 
The relevant steps are outlined in the following.

Using Eq.(\ref{ta}), we find for the initial distance corresponding to 
the travel time $t^1_{g_1}$: 
\begin{eqnarray} \label{x1b}
  x_0 = x^1_{g_1} = \frac{c\,e^1}{H}\left[Ei\left(-\exp(t^1_{g_1}\,H)
  \right) - Ei(-1) \right]\ .
\end{eqnarray}
For the return signal from B to A the initial distance is then
$x^2_{g_1}=x^1_{g_1} \exp(t^1_{g_1}\,H)$, which allows to calculate 
the return time $t^2_{g_1}$ using Eq.(\ref{ta}):
\begin{eqnarray} \label{t2b}
  \exp(t^2_{g_1}\,H)=-Ei^{-1}\,\left(\frac{H\,x^2_{g_1}}{c\,e^1}
  +Ei(-1)\right)\,,
\end{eqnarray}
where $Ei^{-1}$ is the inverse exponential integral. 

Together with $t^1_{g_1}=t_a/2$ any value $t^1_{g_2}>0$ with 
$t_{g_2}=t^1_{g_2}+t^2_{g_2}<t_a$ will give a starting point for the 
bisection method. A suitable choice is $t^1_{g_2}=t^1_{g_1}-(t_{g_1}-
t_a)$, implying $t^1_{g_2}+t^2_{g_1}=t_a$. As $t^1_{g_1}<t^1_{g_1}$ 
implies$t^2_{g_2}<t^2_{g_1}$, we have 
$t_{g_2}=t^1_{g_2}+t^2_{g_2} <t_a$, and the pair $t^1_{g_2}$, 
$t^1_{g_1}$ provides staring values for the bisection method, 
continuing with $t^1_{g_3}=(t^1_{g_2}+t^1_{g_1})/2$, replacing 
in the next step either $t^1_{g_2}$ (for $t_{g_3}<t_a$) or $t^1_{g1}$ 
(for $t_{g_3}>t_a$) with $t^1_{g_3}$, and so on. After completing its 
calculation A sends the result to B, who then sets its clock 
accordingly.

For the numerical example we have $t^1_{g_1}=t_a/2=2.5\times 10^9\ 
[y]$, and get $t_{g_1}=5.5661\times 10^9\ [y]$, $t^1_{g_2}=t^1_{g_1}
-(t_{g_1}-t_a) = 1.93391\times 10^9\ [y]$. With 
$t^1_{g_1}$ and $t^1_{g_2}$ we have starting values for our bisection 
which proceeds with $t^1_{g_3}=(t^1_{g_1}+t^1_{g_2})/2=2.21695\times 
10^9\ [y]\ \Rightarrow\ t_{g_3}=4.86586\time 10^9\ [y]$, then replacing 
$t^1_{g_2}$ by $t^1_{g_3}$ because of $t_{g_3}-t_a<0$, continuing 
with a new $t^1_{g_3}=(t^1_{g_1}+t^1_{g_2})/2$, and so on, till 
convergence is reached at $t_a=t_{g_3}=2.271888\times 10^9\ [y]$. 
A signals this result for $t^1_a$ to B, and B can set its clock 
accordingly. Figure (\ref{fig_AB}) illustrates the exchange of 
signals. There are deviations from straight lines which are not 
visible on the scale of this figure.
\bigskip % \vfil\eject

{\bf 3. Summary  \label{Sum}} \medskip

A simple model of an expanding universe is introduced. It allows for 
a variety of analytical calculation. They are intended for qualitative,
pedagogical value and not of quantitative relevance. Results include: 
\medskip

\noindent
Trajectories of light. See Fig.(\ref{fig_xt}) for examples.
\medskip

\noindent
The initial distance $x_0$ as function of the arrival time $t_a$ (or
vice versa) and the Hubble horizon $x_0^{\max}$ are drawn in Fig.(
\ref{fig_ta}). 
\medskip

\noindent
Synchronization of clocks to a common time is outlined
in section 3, and a diagram of the corresponding exchange of signals
is given in Fig.(\ref{fig_AB}).
\bigskip % \vfil\eject

{\bf Acknowledgments: \label{Ack}} I have not seen them the exercises 
presented here in the literature, but cannot exclude that someone has 
them (or variations of them) worked out before. Please, communicate any 
pertinent information to my e-mail address: bberg@fsu.edu.
%\bigskip

\end{document}